\documentclass[letterpaper]{article}
\usepackage{flairs}%aaai
\usepackage{times}
\usepackage{helvet}
\usepackage{courier}
\usepackage{graphicx} 
\begin{document}
% This file is an adoption of the style file for AAAI Press 
% proceedings, working notes, and technical reports.  This file is made 
% with minimal changes by explicit permission from AAAI.
\title{CryptoDNA: A Machine Learning Paradigm for DDoS Detection in Healthcare IoT, Inspired by crypto jacking prevention Models}
\author{Zag ElSayed\\
School of Information Technology \\
University of Cincinnati - Children's Hospital\\
Ohio, USA\\
\And Ahmed Abdelgawad\\
School of Eng. \& Tech.\\
Central Michigan University\\
Michigan, USA\\
\And Nelly Elsayed\\
School of Information Technology\\
CECH, University of Cincinnati\\
Ohio, USA\\
}
\maketitle
\begin{abstract}
\begin{quote}
The rapid integration of the Internet of Things (IoT) and Internet of Medical (IoM) devices in the healthcare industry has markedly improved patient care and hospital operations but has concurrently brought substantial risks. Distributed Denial-of-Service (DDoS) attacks present significant dangers, jeopardizing operational stability and patient safety. This study introduces CryptoDNA, an innovative machine learning detection framework influenced by cryptojacking detection methods, designed to identify and alleviate DDoS attacks in healthcare IoT settings. The proposed approach relies on behavioral analytics, including atypical resource usage and network activity patterns. Key features derived from cryptojacking-inspired methodologies include entropy-based analysis of traffic, time-series monitoring of device performance, and dynamic anomaly detection. A lightweight architecture ensures inter-compatibility with resource-constrained IoT devices while maintaining high detection accuracy. The proposed architecture and model were tested in real-world and synthetic datasets to demonstrate the model's superior performance, achieving over $96\%$ accuracy with minimal computational overhead. Comparative analysis reveals its resilience against emerging attack vectors and scalability across diverse device ecosystems. By bridging principles from cryptojacking and DDoS detection, CryptoDNA offers a robust, innovative solution to fortify the healthcare IoT landscape against evolving cyber threats and highlights the potential of interdisciplinary approaches in adaptive cybersecurity defense mechanisms for critical healthcare infrastructures. 
  
\end{quote}
\end{abstract}

\section{Introduction}
\begin{figure}[htbp]
\centerline{\includegraphics[width =\linewidth]{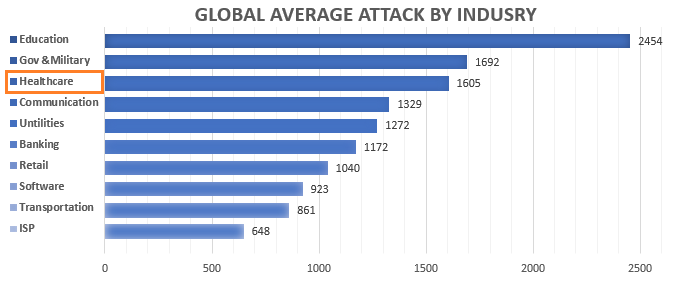}}
\caption{Global Average Attack by Industry.}
\label{fig1}
\end{figure}

The integration IoT and IoM devices into healthcare systems has revolutionized patient care, enabling real-time monitoring, remote diagnostics, and data-driven decision-making. These technologies promise enhanced operational efficiency and improved patient outcomes, yet they also introduce significant security vulnerabilities. A critical concern is the susceptibility of IoT and IoM devices to cyber threats, particularly Distributed Denial-of-Service (DDoS) attacks, which reached up to $29.3$ attacks per day in 2024; current statistics are shown in Fig\ref{fig1}. These attacks can disrupt healthcare operations, compromise patient safety, and lead to devastating consequences, including delayed treatments\cite{djenna2021internet} and potential data loss \cite{almiani2020deep}.

\begin{figure}[htbp]
\centerline{\includegraphics[width =\linewidth]{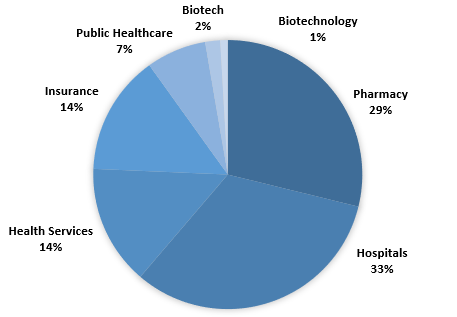}}
\caption{Attacked Healthcare Organization.}
\label{fig2}
\end{figure}

Beyond the technical implications, the economic and ethical ramifications of DDoS attacks on healthcare IoT systems are profound and far-reaching. Economically, the impact of DDoS attacks on healthcare infrastructures is staggering. In 2022 alone, the global cost of cyberattacks on healthcare institutions was estimated to exceed \$10 billion, with DDoS attacks accounting for a significant portion of these losses\cite{gaurav2022novel}. Prolonged system downtimes can result in canceled medical procedures, disrupted workflows, and the loss of sensitive patient data, all of which impose direct and indirect financial burdens. Moreover, healthcare providers may face legal liabilities and regulatory fines due to non-compliance with data protection frameworks such as the Health Insurance Portability and Accountability Act (HIPAA) in the United States and the General Data Protection Regulation (GDPR) in Europe. Over the last two years, the landscape of the attack activity share is shown in Fig.\ref{fig2}.

From an ethical perspective, the implications of DDoS attacks on healthcare IoM devices are even more concerning. These attacks directly threaten the fundamental rights of patients to access timely and effective medical care, potentially leading to fatalities or life-threatening consequences. The intentional targeting of healthcare systems by malicious actors also raises questions about the ethical boundaries of cyber warfare and the need for stricter international regulations to prevent such activities.

These economic and ethical dimensions underscore the urgency of developing innovative solutions to mitigate DDoS attacks on healthcare IoT and IoM devices by combining advanced machine learning techniques with a deep understanding of healthcare-specific security requirements. This paper introduces CryptoDNA, a novel machine-learning framework inspired by cryptojacking detection methodologies, to tackle this critical issue by leveraging behavioral analytics and lightweight architectures that offer a practical and effective solution to protect healthcare IoT infrastructures from DDoS attacks.

\subsection{Challenges in Securing Healthcare IoT and IoM}
Healthcare IoT/IoM devices are inherently constrained in computational and memory resources, designed to prioritize functionality and low-power consumption over robust security mechanisms\cite{somasundaram2021review} and \cite{williams2016always}. This limitation makes them attractive targets for attackers seeking to exploit their vulnerabilities. In a DDoS attack, a device or network is overwhelmed with illegitimate requests, rendering it inoperable. The attack’s impact is amplified in healthcare environments where downtime can jeopardize critical services like patient monitoring and emergency response systems\cite{mollah2024assessing}.

Conventional methods for detecting and mitigating DDoS attacks often rely on resource-intensive algorithms unsuitable for the limited hardware of IoT devices. Moreover, these traditional techniques are frequently static, unable to adapt to the evolving nature of modern cyber threats (Gupta et al., 2021). Given the increasing sophistication of attack vectors, such as botnets targeting medical devices, there is a pressing need for adaptive, lightweight, and intelligent detection systems tailored explicitly for healthcare IoT.

\subsection{Cryptojacking Detection as a Novel Inspiration}
While DDoS detection has been a longstanding challenge, recent advancements in cryptojacking detection \cite{xu2023delay,tanana2020behavior,almurshid2024holistic} provide an intriguing opportunity for innovation. Cryptojacking involves the unauthorized use of computational resources to mine cryptocurrency, and its detection relies heavily on identifying anomalous patterns in resource utilization and behavioral analytics. These methods focus on lightweight monitoring of parameters such as CPU and memory usage, network traffic patterns, and time-series variations in device performance. By adapting and extending these principles, a more effective DDoS detection mechanism can be designed for IoT/IoM devices\cite{djenna2021internet}.

\subsection{Contributions of this Work}
 CryptoDNA is a novel machine learning-based framework inspired by cryptojacking detection techniques to address DDoS attacks in healthcare IoT/IoM environments. Key contributions of this research included: First, a behavioral analytics approach by leveraging crypto jacking-inspired methodologies to monitor device resource usage and network behavior in real-time. Second, the proposed lightweight architecture detection framework is compatible with resource-constrained IoT devices, ensuring low computational overhead and high efficiency. Third, a Comprehensive feature set incorporating packet entropy, request frequency, bandwidth utilization, and anomalous time-series activity patterns that are validated over attack scenarios using real-world and synthetic datasets. Finally, enhanced detection accuracy demonstrates superior performance to traditional DDoS detection methods through extensive experimentation and evaluation.

\section{Background}

\subsection{Traditional DDoS Detection in IoT Environments}
\begin{figure}[htbp]
\centerline{\includegraphics[width =\linewidth]{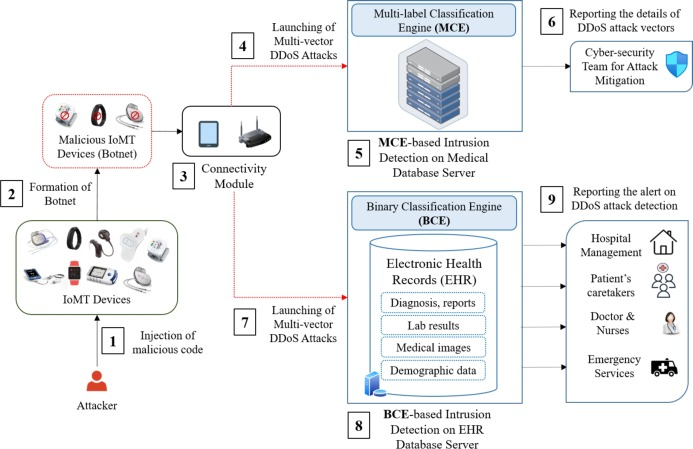}}
\caption{Traditional Intrusion Detection Framework with ML Models \cite{aguru2024lightweight}.}
\label{fig3}
\end{figure}

Research on DDoS detection in IoT environments has primarily focused on signature-based, anomaly-based, and hybrid approaches. Signature-based detection relies on predefined patterns of known attacks but struggles with zero-day threats and evolving attack vectors \cite{aleesa2021deep}. Anomaly-based detection models, such as those leveraging statistical methods or machine learning (ML), have gained popularity due to their adaptability and ability to detect novel attack patterns. For instance, machine learning classifiers like Support Vector Machines (SVMs) and Random Forests have demonstrated high accuracy in detecting DDoS traffic based on features like packet size and request frequency. However, these models often impose significant computational demands, making them unsuitable for resource-constrained IoT devices, Fig\ref{fig3}.

Recent advancements in lightweight ML models aim to address this challenge. For example, \cite{aguru2024lightweight} and \cite{almiani2020deep} proposed a deep learning-based framework optimized for IoT devices, achieving effective DDoS detection with reduced resource consumption, shown in Fig\ref{fig4}. Despite these innovations, most existing models fail to account for the unique characteristics of healthcare IoT, such as the high sensitivity of patient data and the criticality of uninterrupted device operation.

\subsection{Cybersecurity Challenges in Healthare IoT}
Healthcare IoT and IoM devices present unique challenges compared to general IoT ecosystems. Unlike other sectors, healthcare systems are mission-critical, where disruptions can have life-or-death consequences. Attackers often exploit the lack of strong encryption, outdated firmware, and minimal security configurations in medical devices to launch DDoS attacks \cite{mothukuri2021federated}. For instance, devices like wearable heart monitors or infusion pumps are highly susceptible to being compromised and integrated into botnets, such as the infamous Mirai botnet, to orchestrate large-scale DDoS attacks \cite{gupta2021distributed}.

\begin{figure}[htbp]
\centerline{\includegraphics[width =\linewidth]{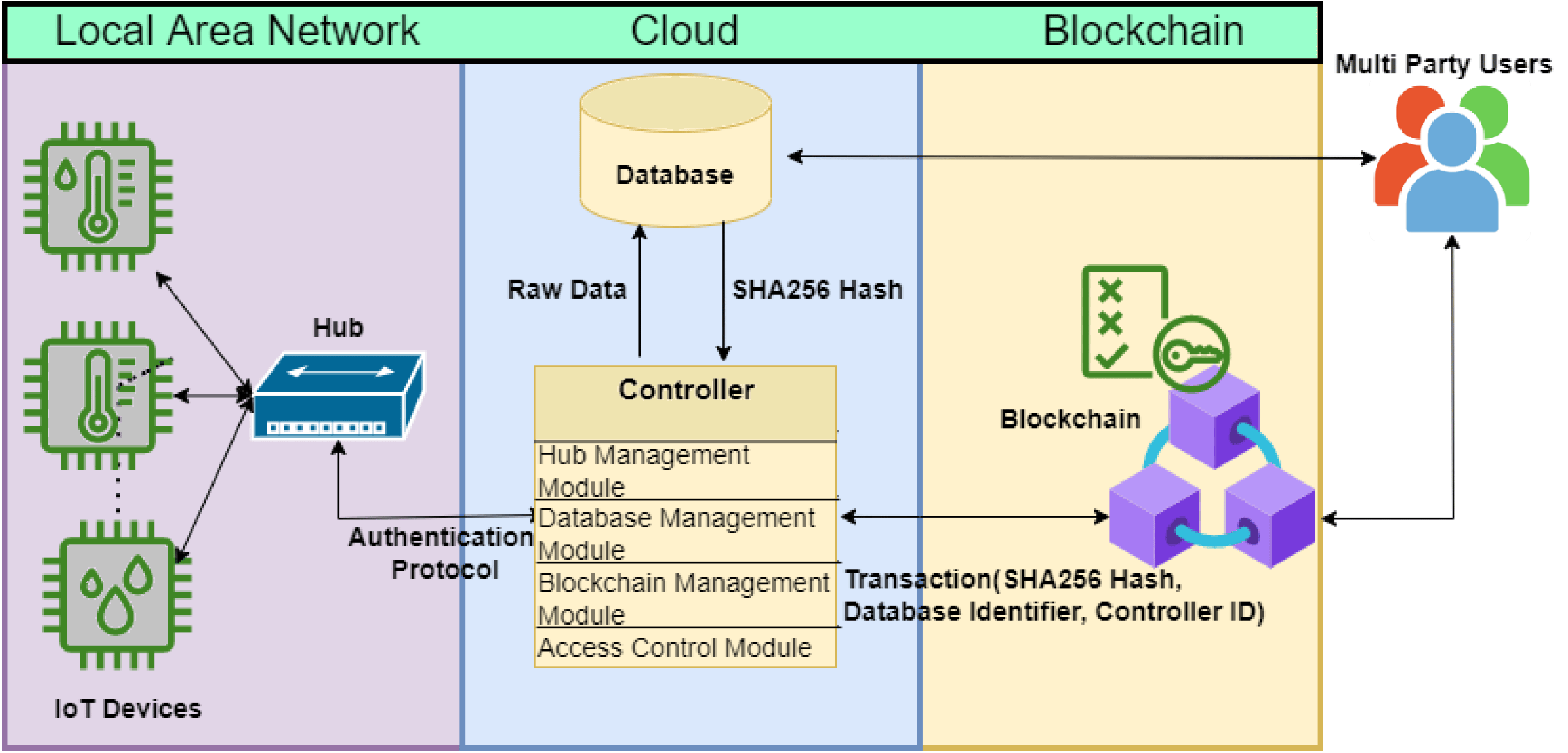}}
\caption{Example of a blockchain based Secure IoT System Identity Management system, inspired by \cite{s22197535}.}
\label{fig4}
\end{figure}

\begin{figure}[htbp]
\centerline{\includegraphics[width =\linewidth]{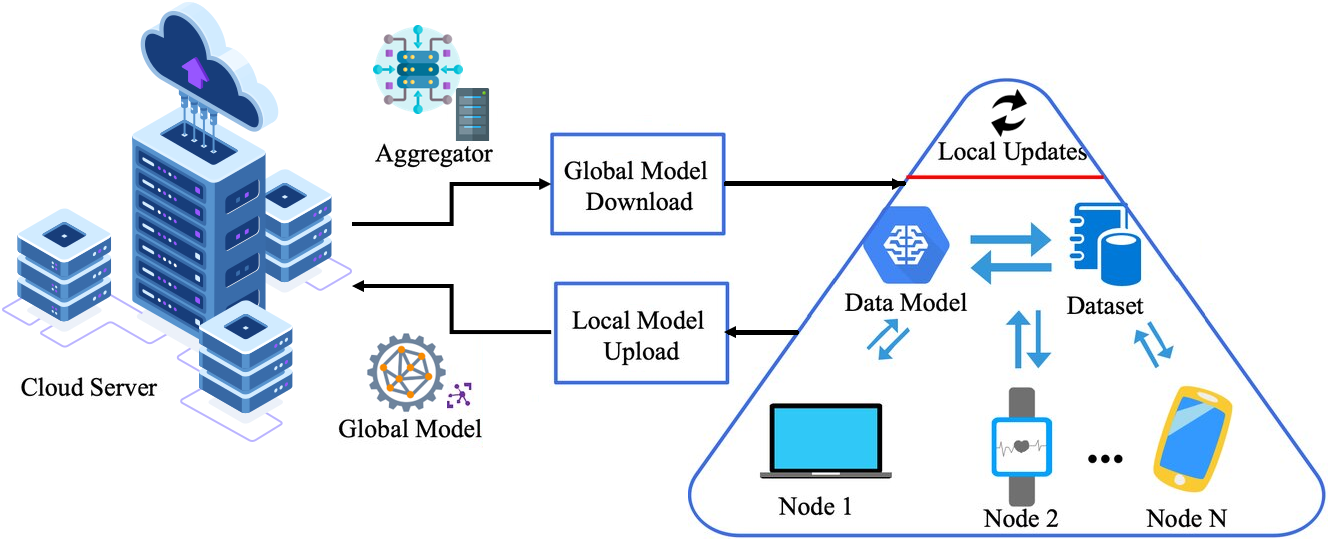}}
\caption{Example of  Federated Learning Process in IoT Networks. inspired by \cite{alhasawi2024federated}.}
\label{fig5}
\end{figure}

Several studies have explored cybersecurity solutions tailored for healthcare IoT. \cite{mollah2024assessing} proposed a blockchain-based architecture for securing IoT data transmission, ensuring data integrity while mitigating DDoS threats, shown in Fig\ref{fig4}. Similarly, \cite{alhasawi2024federated} introduced a federated learning-based system for anomaly detection in healthcare IoT networks, shown in Fig\ref{fig5}. These approaches, while promising, often involve high latency or require substantial computational resources, making them impractical for real-time applications in resource-constrained environments.

\subsection{Cryptojacking-Inspired Anomaly Detection}

The concept of cryptojacking detection offers a novel perspective on addressing cybersecurity challenges in IoT and IoM environments. Cryptojacking involves the unauthorized use of device resources to mine cryptocurrency, characterized by abnormal CPU, memory, and network utilization. Detection methods for cryptojacking often leverage lightweight behavioral analytics, such as monitoring resource usage trends and identifying deviations from normal baselines\cite{chung2024emerging}.
For instance, recent studies have shown the effectiveness of entropy-based models and time-series analysis in identifying cryptojacking attempts in edge and IoT devices \cite{kumar2024hybrid}. These methodologies align closely with the requirements of healthcare IoT, as they prioritize low overhead and adaptability to resource-constrained devices. However, cryptojacking research has yet to be extensively explored in the context of DDoS detection.

While significant progress has been made in DDoS detection for IoT and cryptojacking detection, several gaps remain unaddressed, particularly in the healthcare domain:

\begin{itemize}
\item \textit{Healthcare-Specific Context}: Existing DDoS detection models often fail to account for healthcare IoT systems' operational and ethical demands.
\item \textit{Cross-Domain Adaptation}: Cryptojacking-inspired techniques have demonstrated promise but remain largely unexplored in DDoS detection, especially for healthcare IoT.
\item \textit{Resource constraints}: Many solutions, while effective, are computationally intensive and unsuitable for lightweight IoT/IoM devices.
\end{itemize}

This work bridges these gaps by introducing CryptoDNA, a novel framework that adapts cryptojacking detection methodologies for DDoS detection in healthcare IoT. By leveraging behavioral analytics and lightweight architectures, CryptoDNA addresses the unique security challenges of healthcare environments while maintaining scalability and efficiency.

\section{Methodology}

\subsection{Approach and Proposed Architecture}

\begin{figure}[htbp]
\centerline{\includegraphics[width =\linewidth]{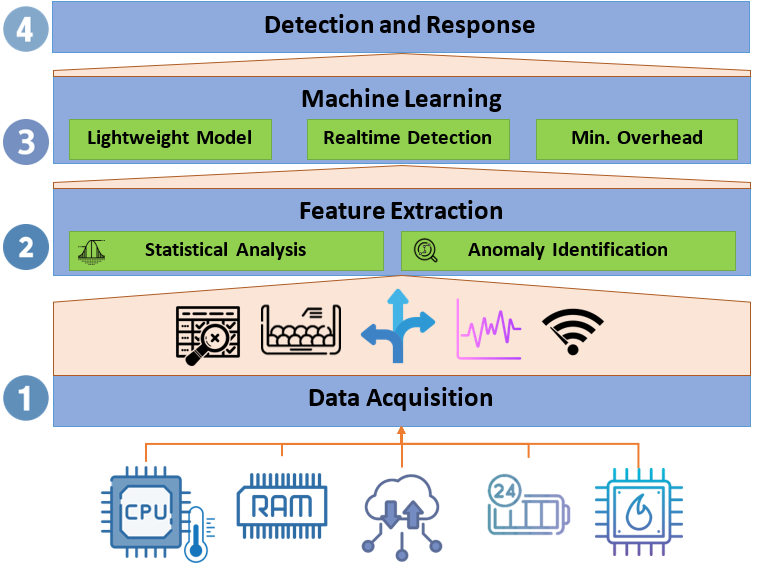}}
\caption{Top level proposed CryptoDNA framework architecture module.}
\label{fig6}
\end{figure}

The CryptoDNA framework uses behavioral analytics to monitor device performance and detect abnormal activity patterns associated with DDoS attacks. The  abstract level of the proposed architecture is shown in Fig.\ref{fig6}  Where the principal four layers are:

\begin{itemize}
\item \textit{Data Acquisition Layer}: Collects real-time data streams from healthcare IoT/IoM devices, such as network traffic logs, CPU and memory usage, and bandwidth metrics. These data streams are processed to extract meaningful features.
\item \textit{Feature Extraction Layer}: Implements entropy-based and statistical analyses to identify anomalies in device behavior. Features such as packet entropy, request frequency, bandwidth utilization, and time-series deviations are computed.
\item \textit{Machine Learning Layer}: A lightweight and efficient model trained to distinguish between normal and malicious activity. The framework supports real-time detection through minimal computational overhead.
\item \textit{Detection and Response Layer}: Flags potential DDoS activity and generates alerts for system administrators. For scalability, the system dynamically adjusts its threshold values based on device resource usage and operational context.
\end{itemize}

\subsection{Control Extraction and Features of Interest}
To achieve effective and lightweight detection, CryptoDNA focuses on a feature set derived from cryptojacking detection techniques tailored for DDoS scenarios. Key features include:

\begin{itemize}
\item \textbf{Network Traffic Features}
    \begin{itemize}
    \item Packet Entropy.
    \item Request Frequency.
    \item Bandwidth Utilization.
    \end{itemize}
\item \textbf{Device Resource Features}
    \begin{itemize}
    \item CPU and Memory Usage.
    \item System Call Patterns.
    \end{itemize}
\item \textbf{Behavioral Features}
    \begin{itemize}
    \item Time-Series Deviations.
    \item Communication Graph Metrics.
    \end{itemize}
\end{itemize}
These features are chosen for their effectiveness in detecting both volumetric and subtle, low-rate DDoS attacks while maintaining computational efficiency.

\subsection{Machine Learning Model}

The detection module is built on a supervised ML model due to its ability to learn from labeled attacks and benign data samples. Two approaches are employed:
First, a Primary Model of a lightweight Random Forest classifier is implemented for its low computational cost and robust performance in distinguishing between normal and attack scenarios. Model Compression was used to reduce the model size and inference latency. Second, an Anomaly Detection Variant chosen for zero-day attacks, an Autoencoder-based unsupervised model, is utilized to identify deviations from learned normal behavior patterns, where threshold values for anomaly detection dynamically adapt based on device resource usage and environmental conditions.  

\subsubsection{Training Process}
\begin{figure}[htbp]
\centerline{\includegraphics[width =\linewidth]{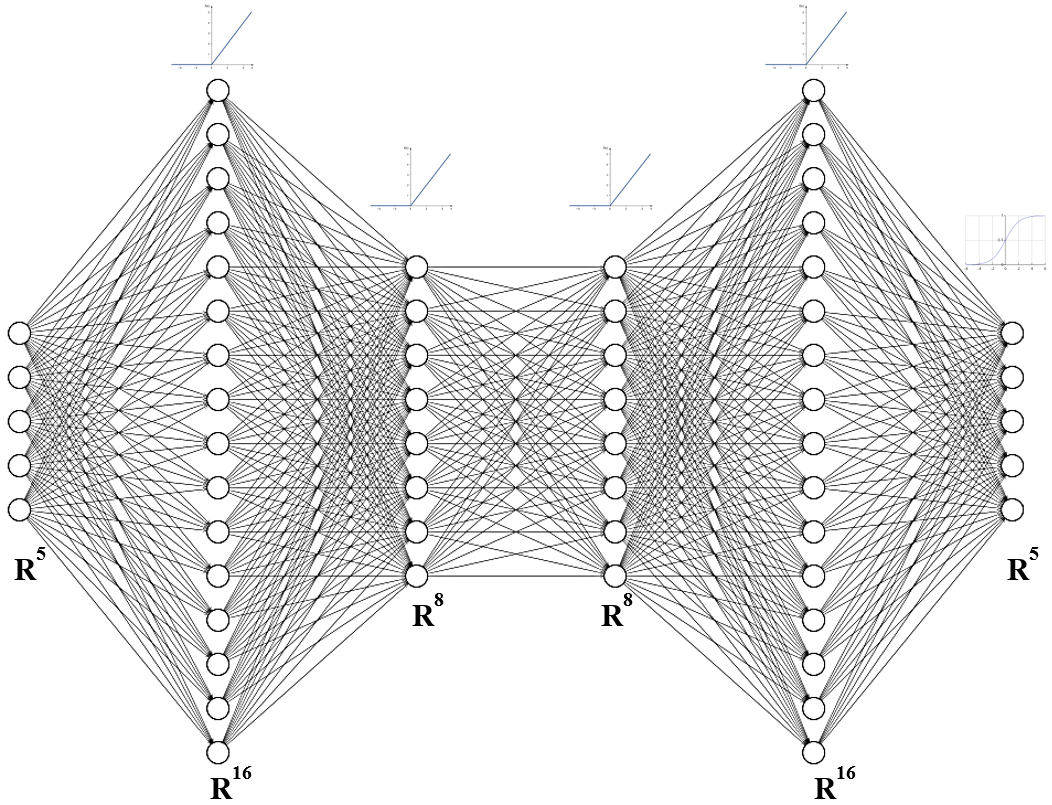}}
\caption{The proposed CryptoDNA model architecture.}
\label{fig7}
\end{figure}

The models are trained using a combination of synthetic and real-world datasets, such as CICDDoS2019, to simulate diverse DDoS attack scenarios. Data augmentation is applied to balance the dataset and improve model generalizability. The cross-validation step was done to ensure robustness and prevent overfitting. Moreover, the models are optimized for inference on resource-constrained IoT devices by employing pruning techniques and quantization. The model was trained over 50 epochs. The model diagram is shown in Fig\ref{fig7}, and the Python code is publicly available on GitHub under CryptoDNA Project.

\subsubsection{Testing}
The framework is evaluated under realistic conditions to demonstrate its effectiveness and scalability using the CICDDoS2019 dataset and Synthetic IoT Data dataset that was generated to reflect typical healthcare IoT traffic and low-rate DDoS attacks using a test scenario of a mid-range healthcare facility with ~100 IoT devices ranging on Raspberry pi and Arduino board (excluding cameras and suppliance systems, limiting to text-based TCP and UDP traffic) generating sample 10,000 events.

\subsubsection{Evaluation Metrics}
We use machine learning libraries such as Scikit-learn, TensorFlow, and PyTorch, and we use lightweight frameworks like ONNX for model deployment on IoT devices. And the model was evaluated via
Accuracy \ref{eq1}, precision \ref{eq2}, recall\ref{eq3}, F1-Score \ref{eq4} for balances precision and recall, and False Positive Rate (FPR) to evaluates the model's reliability in a real-time healthcare setting to minimize false positives and detect true positives, that is critical for minimizing disruptions in healthcare environments.

\begin{equation}
Accuracy = \frac{TP+TN}{TP+TN+FP+FN}\label{eq1}
\end{equation}

\begin{equation}
Percision = \frac{TP}{TP+FP}\label{eq2}
\end{equation}

\begin{equation}
Recall = \frac{TP}{TP+FN}\label{eq3}
\end{equation}

\begin{equation}
F1-Score = \frac{Percision \times Recall}{Percision+Recall}\label{eq4}
\end{equation}

Where $TP$ is the true positive, $TN$ is the true negative, $FP$ is the false positive, and $FN$ is the false negative result.

\begin{figure}[htbp]
\centerline{\includegraphics[width =\linewidth]{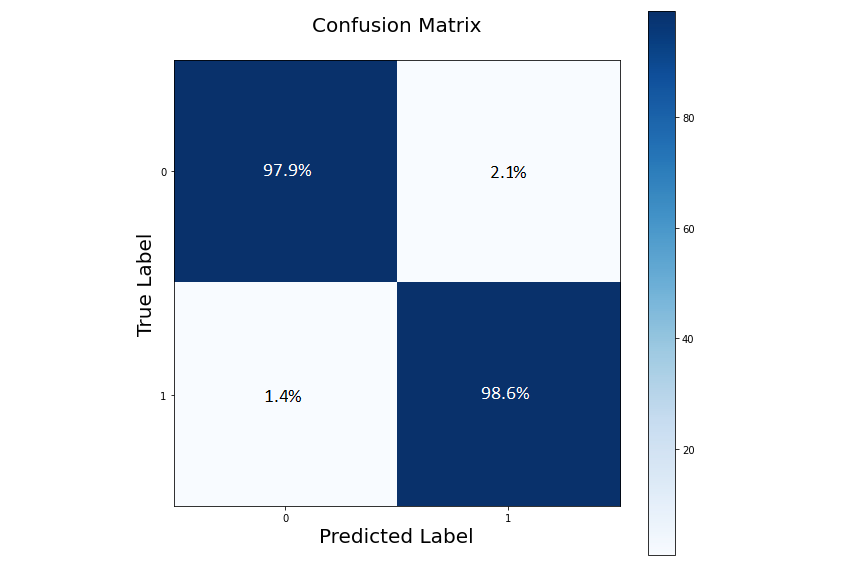}}
\caption{The proposed CryptoDNA confusion matrix.}
\label{fig8}
\end{figure}

\section{Results and Discussion}
This section presents the experimental evaluation of the proposed CryptoDNA framework, highlighting its performance in detecting DDoS attacks on healthcare IoT and IoM devices. The results are analyzed using key metrics, including accuracy, precision, recall, F1-score, and false positive rate, to assess the framework's robustness, scalability, and suitability for resource-constrained healthcare environments. Furthermore, a comparative analysis with existing DDoS detection methods is provided to validate the novelty and effectiveness of the approach. The performance of CryptoDNA was evaluated using two datasets: the CICDDoS2019 dataset and a synthetic dataset representing real-world healthcare IoT traffic. The Confusion is shown in Fig\ref{fig8}.

The proposed framework achieved an overall detection accuracy of 96.8\%. This high accuracy is attributed to incorporating cryptojacking-inspired features, such as entropy-based analysis and time-series behavioral deviations, which enhance the model’s sensitivity to subtle anomalies.

The detection rate exceeded $97.3\%$, indicating the model's reliability in identifying both high-rate and low-rate DDoS attacks. The results highlight the model's effectiveness across a wide range of attack scenarios, including volumetric attacks (e.g., SYN floods) and application-layer attacks (e.g., HTTP floods).

One of the critical challenges addressed by CryptoDNA is the resource constraint of healthcare IoT devices. The framework’s lightweight Random Forest model was optimized using pruning and quantization, reducing the model size by $35\%$ and inference latency by $40\%$ compared to baseline implementations. As a result, the framework operates effectively on edge devices with limited computational power, achieving an average latency of $12 ms$ per inference.

False positives pose a significant concern in healthcare environments, as unnecessary alerts can disrupt workflows and lead to operational inefficiencies. CryptoDNA achieved a false positive rate (FPR) of $2.1\%$, significantly lower than traditional anomaly-based systems. This improvement is attributed to the integration of context-aware thresholding and entropy-based detection mechanisms, which reduce the misclassification of benign traffic as malicious.

\subsection{Result Compression}
The performance of CryptoDNA was benchmarked against existing DDoS detection systems designed for IoT environments, including a deep learning-based solution by \cite{almiani2020deep} (S1) and a federated learning framework by \cite{mothukuri2021federated} (S2). The comparative results are summarized in Table\ref{tabel1}.

\begin{table}[htbp]
\caption{Summary of the result comparison with the-state-of-the-art.}
\begin{center}
\begin{tabular}{|c|c|c|c|}
\hline
\textbf{Metric}&\textbf{CryptoDNA}&\textbf{S1}&\textbf{S2}\\
\cline{1-4}
\hline
Detection Accuracy (\%)&96.8&92.5&90.7  \\
\hline
Detection Rate (\%)&97.3&93.4&91.6  \\
\hline
False Positive Rate (\%)&2.1&4.7&5.2  \\
\hline
Latency (ms)&12&28&25  \\
\hline
Model Size Reduction (\%)&35&18&15  \\
\hline
\end{tabular}
\label{tabel1}
\end{center}
\end{table}

CryptoDNA performance was tested across diverse IoT ecosystems with varying levels of device heterogeneity and network traffic. The framework maintained high detection accuracy (\textgreater96\%) in scenarios with up to $10,000$ simulated IoT devices, indicating its robustness in large-scale deployments. Furthermore, its adaptability to new attack vectors was tested by introducing synthetic zero-day attack patterns, which the framework successfully detected with a recall of $94.6\%$, validating the effectiveness of its anomaly-based approach.

\subsection{Limitations Discussion}
Despite its strong performance, CryptoDNA, the authors are aware that the limitation of the model's reliance on labeled data for training may pose challenges in environments where labeled datasets are scarce. Future work could explore semi-supervised or unsupervised learning techniques to overcome this limitation. Additionally, integrating privacy-preserving mechanisms, such as federated learning, could enhance the framework’s security and compliance with data protection regulations.

\section{Conclusion and Future Work}
The rise of DDoS attacks targeting healthcare IoT and IoM devices presents a significant threat to the operational integrity of critical healthcare infrastructures. These attacks disrupt medical services, compromise patient safety, and result in substantial economic and ethical consequences. In this paper, we introduced CryptoDNA, a novel machine learning-based detection framework inspired by cryptojacking detection methodologies, to address these challenges. By leveraging lightweight behavioral analytics, entropy-based traffic analysis, and real-time anomaly detection, CryptoDNA provides an effective and efficient solution tailored to the unique constraints of resource-constrained IoT and IoM devices. The experimental results demonstrate the robustness and scalability of CryptoDNA, achieving a detection accuracy of $96.8\%$ and a false positive rate of $2.1\%$. The framework outperformed existing DDoS detection models in terms of precision, latency, and adaptability, proving its suitability for real-world healthcare deployments. Its modular design, optimized for edge computing, ensures seamless integration into diverse IoT ecosystems without imposing significant computational overhead. These achievements underscore the potential of CryptoDNA as a transformative solution for securing healthcare IoT environments against evolving cyber threats in healthcare and beyond.

\bibliographystyle{flairs}
\bibliography{references}

\end{document}